*Research Article*

# Anchor-Less Producer Mobility Management in Named Data Networking for Real-Time Multimedia

**Inayat Ali [1,2] and Huhnkuk Lim [1,2]**

[1] *Korea Institute of Science and Technology Information (KISTI), Daejeon, Republic of Korea*
[2] *University of Science and Technology (UST), Daejeon, Republic of Korea*

Correspondence should be addressed to Huhnkuk Lim; hklim@kisti.re.kr

Information-centric networking (ICN) is one of the promising solutions that cater to the challenges of IP-based networking. ICN shifts the IP-based access model to a data-centric model. Named Data Networking (NDN) is a flexible ICN architecture, which is based on content distribution considering data as the core entity rather than IP-based hosts. User-generated mobile contents for real-time multimedia communication such as Internet telephony are very common these days and are increasing both in quality and quantity. In NDN, producer mobility is one of the challenging problems to support uninterrupted real-time multimedia communication and needs to be resolved for the adoption of NDN as future Internet architecture. We assert that mobile node's future location prediction can aid in designing efficient anchor-less mobility management techniques. In this article, we show how location prediction techniques can be used to provide an anchor-less mobility management solution in order to ensure seamless handover of the producer during real-time multimedia communication. The results indicate that with a low level of location prediction accuracy, our proposed methodology still profoundly reduces the total handover latency and round trip time without creating network overhead.

## 1. Introduction

Mobile devices have transformed into multimedia computers owing to the multitude of mobile applications and diverse features, such as cameras, messaging, online content sharing, and online mobile gaming. These smart features of mobile devices and the human needs of creating and sharing contents in real-time have enabled mobile devices to play the roles of content providers and content consumers at the same time. The exponential growth in data production and its need for dissemination is becoming a cumbersome issue in current network architecture. The existing Internet model is not built to combat this growing production and dissemination of data. Moreover, many attempts were made to withstand this issue, which resulted in peer-to-peer (P2P) overlays on the IP network and content distribution network (CDN) [1]. These attempts are still very feeble to ensure smooth network operation. According to a prediction by CISCO Visual Networking Index (VNI) [2], global traffic will inflate to 3.3 ZB per year, or 278 Exabytes per month by 2021, thus raising more disputes for the current state of the practice network architecture and its protocols.

Many proposals have been witnessed in recent years for a scalable and reliable future Internet architecture [3–6]. Among these proposals, information-centric networking (ICN) has gained much attention. ICN is a data-oriented network, where the information is retrieved from the Internet by naming the data, not the end hosts. Data objects are independent of location unlike the IP address in the traditional network. This novel Internet architecture has addressed the challenges faced by the current Internet architecture including scalability, addressing, name resolution, security, privacy, routing, and mobility. Among the ICN approaches, Named Data Networking (NDN) [7] is a very active, agile, and enterprising one. NDN operation is based on Interest/Data packets, where a consumer (data requester) sends an Interest packet containing the name of the required data. The NDN Forwarding Information Base (FIB) helps forward the Interest packet towards the data named in the Interest packet and retrieve the data in one or more Data packets. NDN also uses in-network caching for faster data access; hence the data can be retrieved from the producer or the cache of the NDN router in the network. Many research challenges in NDN including naming, name resolution, routing, security for large-scale data, and mobility need to be resolved for NDN to win the race of future Internet architecture. The number of mobile devices and the data they generated have drastically increased over time [8, 9]. Mobility management is becoming more salient because of the device transformation into a smart gadget that rapidly creates and shares content in real time. Therefore, device mobility management needs much attention from the research community. As a result of NDN design principles, consumer mobility is natively supported by NDN as a change in the physical location of the consumer does not affect the NDN data plane. There is no need for signalling from network to heal from consumer mobility but only Interest retransmission by the consumer for the data that have not yet been received works well. Handling producer mobility is a challenging task and needs to update the network data plane to successfully recover from outages due to producer mobility.

A few producer mobility management schemes for ICN have been proposed for real-time services. However, there is still room to design sophisticated algorithms to handle producer mobility efficiently. The work in [10, 11, 12] has reactive approaches towards mobility management where the packets are redirected after the handover completes and notification is received from the producer. These approaches suffer from path stretch problem like in Mobile IP for mobility management in IP networking. Moreover, these reactive approaches encounter more content retrieval delay than proactive approaches. The concept of redirecting packets toward the new location was first introduced in the traditional IP network (Mobile IP [13]) to support mobility. However, in mobile IP, all the packets have to be redirected by the old access point towards the new location until the communication session ends, thus causing path stretch problem for the entire session. The techniques used in [14, 15] use a proactive caching approach to minimize the Interest retransmission during a handover. The scheme in [14] assumes that the request pattern is known and the future requested content is proactively pushed to the network caches ahead of handover to satisfy the Interest during the handover. However, the mechanism in [15] pushes the content to network cache based on content popularity. However, these approaches cannot be applied in real-time communication, where the contents are produced in real-time after the Interest packets are received. Proactive caching techniques assume that the data are already extant and pushed to the network cache before receiving an Interest for those data, which is not the case in real-time communication. Real-time data are the data that are generated and delivered immediately after they are requested. The data generated in Internet telephony, messaging, and online gaming are examples of real-time data, and the above proactive caching techniques for mobility management will not support these applications, which account for a significant amount of Internet traffic today.

To solve the problems associated with the above techniques, i.e., path stretch and long delay due to reactive approaches and lack of support for real-time communication in proactive caching approaches, we, in this article, propose a proactive mechanism that is the first attempt to exploit location prediction techniques for anchor-less producer mobility management in order to ensure seamless handover of the producer during real-time multimedia communication. The proposed methodology is based on the prediction of the future location of a moving producer and proactively redirecting the Interests to the new access point (nAP) before the handover completes. The proposed methodology is the first to exploit location prediction techniques in NDN for mobility management. Our scheme reduces both the handover latency and the round trip time without causing network overhead. The mechanism also does not suffer from the path stretch problem. Moreover, it does not assume that the data are already extant; instead, the data are generated in real-time and forwarded after the Interest packets are intelligently delivered to the producer after the handover.

The rest of the paper is structured as follows: In Section 2, we briefly offer an insight into NDN mobility and related work. In Section 3, we describe the proposed location prediction-based mobility management methodology in details. We provide a detailed discussion of the results and comparison with two legacy schemes in Section 4. Finally, we conclude the paper in Section 5 and present future work.

## 2. Insight into NDN Mobility

In this section, we give an overview of operational aspects of NDN, provide problems regarding mobility, and present some related work on mobility management in NDN.

*2.1. Mobility in NDN.* NDN provides a new communication paradigm that revolves around the content/data. The data are retrieved by data names rather than the IP addresses of the machines that host the data. Once the data are forwarded to the network, they are cached at network routers (in-network caching) to satisfy future requests for the same data with small delays. This in-network caching results in faster service access and resists congestion. Two data structures are used by NDN routers to aid packets forwarding, i.e., Forwarding Information Base (FIB) and Pending Interest Table (PIT). FIB is populated by routing protocols and different forwarding strategies, and it is used to forward Interest packets towards the data. PIT keeps the records of interfaces on which the Interest packets are received and the data names requested by those Interests. PIT ensures that the Data packets follow the reverse path of the Interest packets.

A third data structure called Content Store (CS) is used at routers to cache data for satisfying future Interests. Upon receiving an Interest in the router if the requested data are available in the router cache, they will satisfy the Interest without forwarding it to the next hop based on FIB. The Interest packet is forwarded to the next hop based on FIB entry if the data are not available in router cache. Before forwarding the Interest packets, its incoming interface and data name is stored in the PIT for data forwarding. When a Data packet is received at a router, it will be forwarded to the interface in the PIT entry against this data name. The data will be dropped if there is no entry against this data name in the PIT. The operation in the NDN router is explained in Figure 1.

Mobility is an essential feature of future Internet technology. NDN supports mobility and reduces the complexities of mobility support in the IP-based network. Mobility can be of two types, i.e., consumer mobility and producer mobility. NDN naturally supports consumer mobility by Interest retransmission and in-network caching. If a consumer after requesting data moves to another network, the response will be delivered to the router in the previous location, which will be cached there. Consumer at the new network will retransmit the Interest for the same data. In which case, the intermediate routers will immediately respond to the Interest. In case the data are not available at any of the network caches, the Interest will hit the data producer again, and the data will be retrieved with a longer delay. Unlike consumer mobility, producer mobility is a complicated problem. This type of mobility highly affects communication. If a producer moves to another network, the Interests will not be delivered to it unless network converges. This type of mobility cannot be tackled with Interest retransmission, as the new name prefix of the producer will not be known to any NDN router unless network converges. The communication session halts for at least the sum of handover time and the network reconvergence time as shown in Figure 2. The Interest packets are discarded at the producer's old point of attachment (oAP) (11.9 sec in Figure 2) and start delivering again to the producer at the new location after the routing update period expires and the whole network converges (13 sec in Figure 2). To handle this mobility management issues, few techniques have been proposed to ensure seamless handover of producers during real-time streaming. Mobility management algorithms are designed to guarantee seamless handover from an old Access Point (oAP) to a new Access Point (nAP) and should have the following characteristics:

(i) The time required to divert the route to nAP from oAP should be minimum. In other words, the handover latency should be the least possible

(ii) The Interest and content loss during the handover time should also be minimum

*2.2. Related Work.* NDN relies on Interest retransmission to heal from network outages owing to mobility. However, Interest retransmission is not a scalable approach when there are a large number of mobility events; also the end-to-end latency due to mobility in this approach is very high. Therefore, some techniques have been proposed so far to provide more robust and efficient mobility management schemes. The schemes in the literature can be divided into three categories: rendezvous-based schemes are more like DNS system in the current IP-based networking, where a consumer sends a query to a particular node called rendezvous server to find the location of the producer before sending an Interest. The rendezvous server keeps track of the location of all nodes in the network. These techniques are easy to implement, but the drawback is that they cause extra-overhead to maintain the location information of mobile nodes. The rendezvous-based schemes need the early binding of names that affect the NDN data naming. Anchor-based approaches used an anchor node that redirects the Interests to the producer at the new location. The anchor node should always be kept informed of the producer's movement. These schemes possess a single relay point (single point of failure) and suffer from the path stretch problem. Performance evaluation of anchor and anchor-less approaches are given in detail in [10]. The anchor-less approach has no pre-specified node to handle mobility. This approach is hard to implement, but the schemes following such approaches are very efficient in terms of performance. The schemes in [16, 17] are anchor-based mobility management schemes. The schemes proposed in [10, 11] are anchor-less mobility management schemes and suffer from the inherited problems from their corresponding approaches. The proposed methodology in this work is based on the prediction of the future location of a mobile node. These location prediction techniques can be exploited to provide anchor-less mobility management that highly reduces the handover latency and round trip time even with low prediction accuracy.

## 3. Location Prediction-Based Mobility Management

Producer mobility causes degradation in the quality of service during real-time streaming. The services are disrupted when a producer moves and goes out of the radio range of its oAP and connects to another access point (nAP). The Interest packets are dropped at the oAP as the new hierarchical name of the producer is not updated in the network information. The communication session halts until the routing update period expires and the whole network converges with this new producer name information as shown in Figure 2. In NDN, the Interest packets are forwarded using FIB entries that are being populated by routing protocols, while the Data packets follow the traces of Interest packet stored in the PIT to reach the consumer. Hence, the producer is not able to deliver Data packets of its ongoing streaming from its new location right after the handover completes, unless it receives an Interest packet at the new location using its new hierarchical name prefix. This drastically

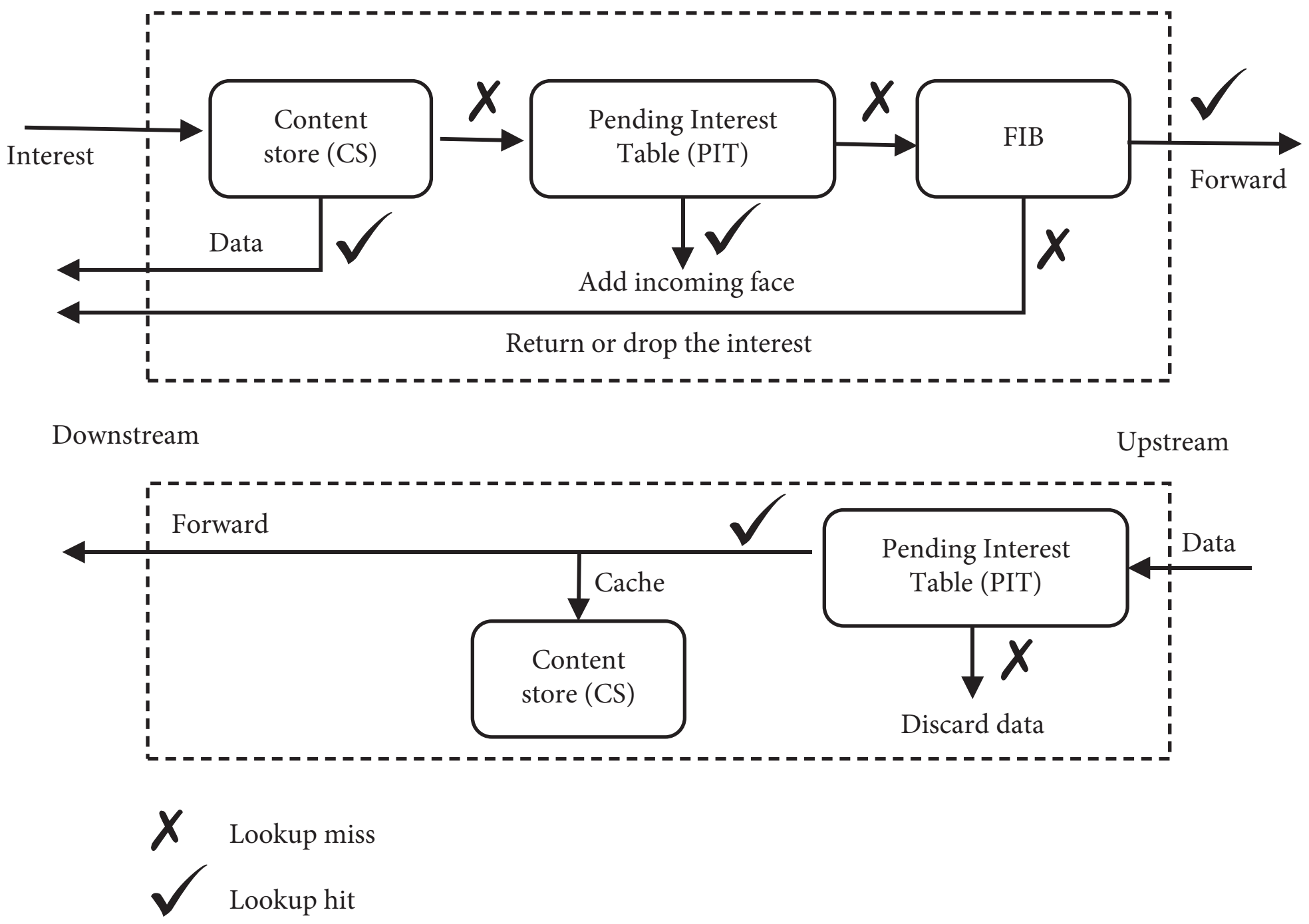

Figure 1: Interest and Data forwarding at the NDN router.

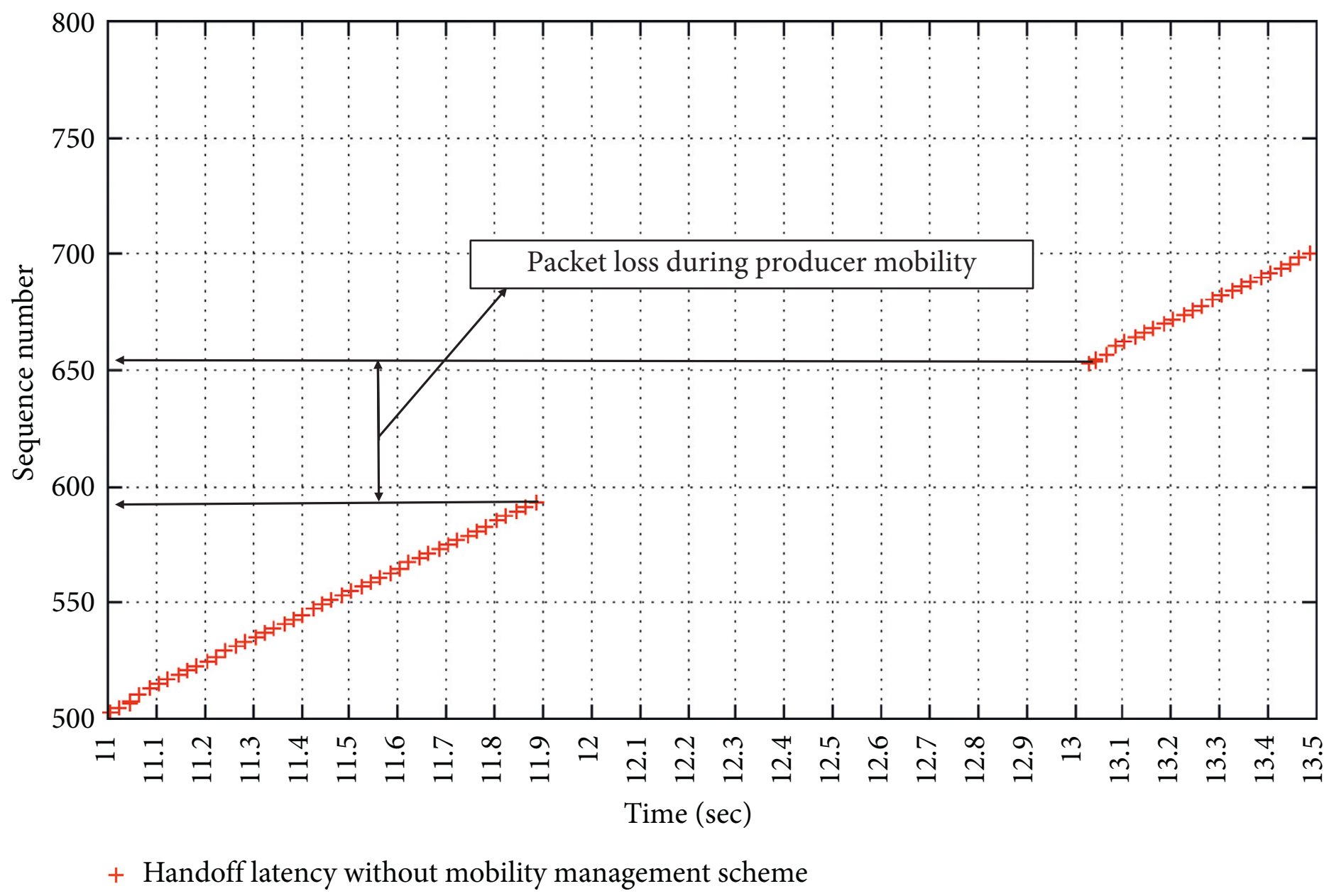

Figure 2: Handover latency without using any mobility management.

maximizes the handover latency and hence round-trip time (RTT).

*3.1. Preliminaries.* We use two types of Interest packets to implement the mobility management scheme efficiently. The Interest types are as follows:

(i) INTEREST_PU: Interest path update message is sent by a mobile node (producer) to its AP to notify it about its mobility and probable handover. The packet format is shown in Figure 3(a).

(ii) INTEREST_RED: Interest redirection is an Interest type in which the Interests are forwarded to the producer at new location after the handover. The packet format is shown in Figure 3(b).

INTEREST_RED packet changes the content name from .../oAP/ale1 to .../nAP/ale1 (since the oAP have predicted the nAP) and is forwarded based on the

| Name | X-coordinates (new predicted position) | Y-coordinates (new predicted position) | Nonce | Interest_PU (value = 0) |
|---|---|---|---|---|

(a)

| Name | Selector (order preference, publisher filter etc..) | Nonce | Guiders (scope, interest life time) | Interest_RED (value = 1) |
|---|---|---|---|---|

(b)

Figure 3: Interest packet types. (a) INTEREST_PU. (b) INTEREST_RED.

FIB entry corresponding to this new name. When an INTEREST_RED packet is received at NDN router (e.g., Rtr-2 in Figure 4), it also updates the name and out going face in the previously recorded PIT entry for content delivery toward the consumer. It is also a very frugal way to use INTEREST_PU message to notify oAP about a mobility event.

*3.2. Algorithm Details.* The mobility management can be controlled at the network or mobile node. However, a more easy and efficient way is to make use of both to develop more efficient mobility management techniques. The proposed methodology manages mobility using cooperation from both the network and the mobile node (producer). The location prediction scheme comes into action when the received signal strength (RSS) drops below the threshold (th = −77 dBm). As the RSS drops below the th, the future location of the producer is calculated by the producer itself using the speed and direction of movement obtained from its sensors. The producer sends its new location to the oAP in an INTEREST_PU message as shown in Figure 4 (Step 3). The INTEREST_PU message contains the $X$ and $Y$ coordinates of the producer's future location. The pseudocode for this is given in Figure 5. The details of our prediction model are given in the following subsection.

*3.2.1. Location Prediction and New AP Selection.* The selection of nAP is the very crucial and important part of this scheme. There are many possible ways to select nAP; however, we use the relative distances of the producer's predicted future location from access points. The relative distance of the producer from APs is used in this work because, in our simulations, all the APs and mobile nodes use the same transmit and receive power. Moreover, we used the free-space propagation model where there are no path losses, and it assumes one clear line-of-sight path from the transmitter to the receiver. In such settings, the relative distance is the fair method to be considered and increase the probability of selecting the right new AP. Upon receiving the INTEREST_PU message from the producer, the oAP finds the nAP using the Euclidean distance formula as explained in the pseudocode in Figure 6. Here, we assumed that all the fixed routers and access points know each other geographic location. To find the distance between producer and nAP at a future time ($T_f$), oAP should know the new location of the producer at $T_f$. There are many methods to predict the future location of a mobile node [18, 19, 20] with different prediction accuracy ranging from 45% to 90%. Our prediction model used in this paper is straightforward where to predict the location at a time ($T_f$), the producer finds its speed and direction of movement and calculates its future location coordinates at the time ($T_f$). The producer sends its new predicted coordinates to oAP in an INTEREST_PU message. The modern mobile devices come with lots of advanced hardware and software sensors like accelerometer, gyroscope, digital compass, GPS, and speedometer. These sensors can easily calculate the speed and direction of movement of the mobile device (producer). The producer constantly checks its RSS after it starts a movement. The connection becomes unreliable after the signal strength drops than a predefined threshold (here th = −77 dBm) and the node starts searching for reliable signals from other APs for handover. Before the handover, the producer gets its speed and direction from the sensors at that particular time and calculates its future coordinates and sends them in an INTEREST_PU message to the oAP. The new speed at a particular time is determined by

$$V_{\text{new}} = \begin{cases} \min\{\max[v_{\text{old}} + \Delta v, 0], v_{\max}\}, & \text{if } p \le p_s, \\ 0, & \text{otherwise}, \end{cases} \quad (1)$$

where $v_{\text{old}}$ is the old speed obtained from the producer device sensors and $\Delta v$ is the change in speed after time ($t$), $\Delta v = v_{\text{new}} - v_{\text{old}}$, and it is uniformly distributed on [−3, 3] km/hr. $v_{\max}$ is the prespecified maximum speed (30 km/hr in our simulation) of a mobile node. $p$ is the uniformly distributed probability between [0, 1], and $p_s$ is threshold probability ($p_s = 0.5$), which determines the probability of producer movement. The direction of the movement is also uniformly distributed and is given by

$$\phi_{\text{new}} = \phi_{\text{old}} + \Delta\phi, \quad (2)$$

where $\phi_{\text{new}}$ is the new direction after time ($t$) and $\phi_{\text{old}}$ is the old direction. $\Delta\phi$ is the change in the direction and is uniformly distributed on $[-\pi/4, \pi/4]$. Euclidean distance between producer and AP is an efficient method to select

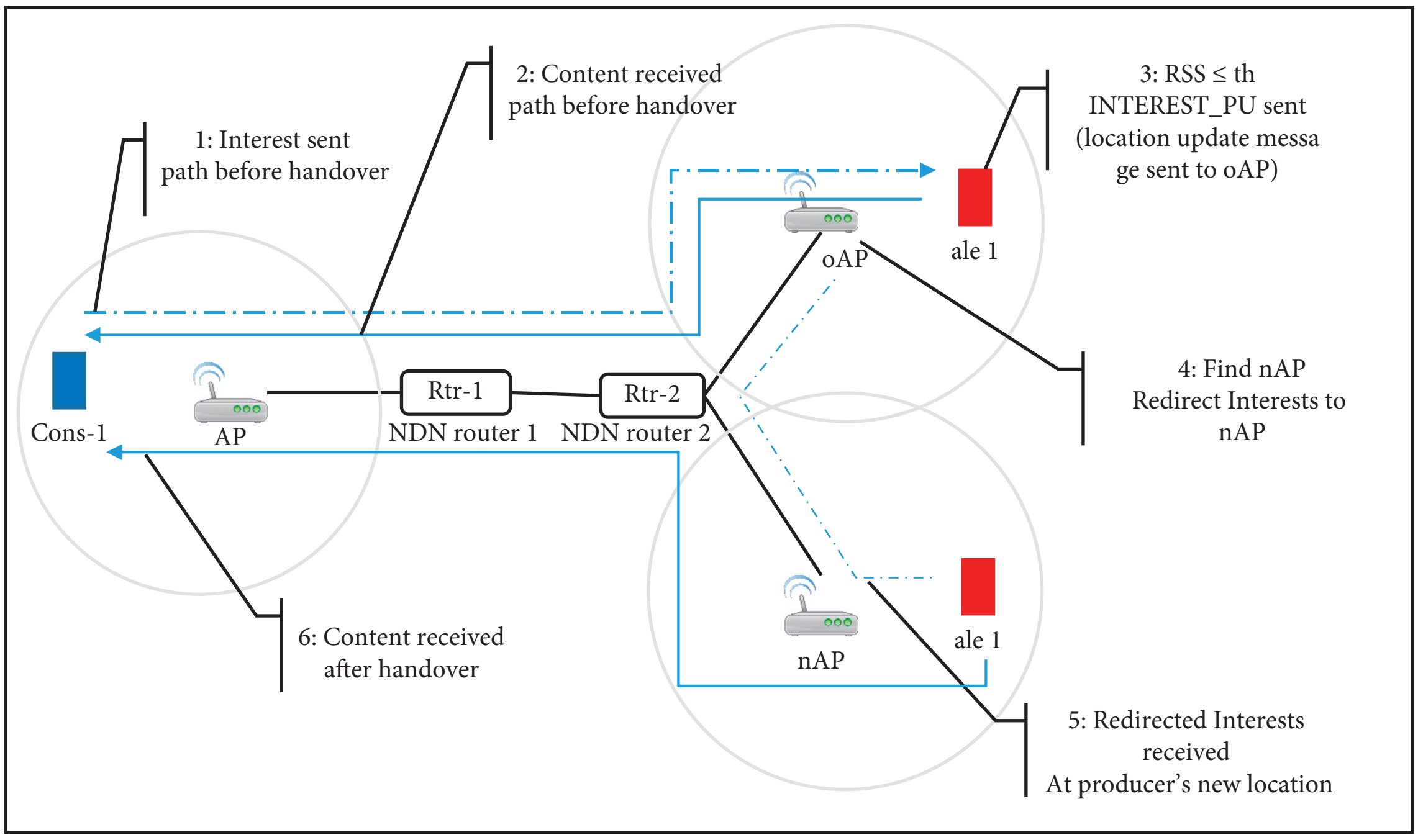

Figure 4: Location prediction-based mobility management scheme.

Finding and sending future location coordinates to oAP

Possible event: Received signal strength (RSS) drops than threshold (th = –77 dBm).

(1) Case event
(2) If RSS ≤ th:
(3) Get speed ($v_{new}$) and direction ($\phi_{new}$)
(4) Calculate future coordinates ($x_p$, $y_p$) using current speed and direction
(5) Construct INTEREST_PU message
(6) Send INTEREST_PU to oAP

Figure 5: Pseudocode for path update message generation.

New AP (nAP) Selection by old AP (oAP)

Possible events: $\zeta$ = Arrival of INTEREST_PU message at oAP, $\partial$ = arrival of nAP new reachable prefix notification from producer at oAP (wrong prediction).

(1) Case events
(2) When $\zeta$:
(3) Get predicted coordinates ($x_p$, $y_p$) from INTEREST_PU message
(4) Predict nAP:
(5) $\text{Dist}_i = \sqrt{(\text{AP}_{i,x} - x_p)^2 + (\text{AP}_{i,y} - y_p)^2}$
(6) $\text{AP}_i = \min(\text{Dist}_i)$
(7) $\text{nAP} = \text{AP}_i$
(8) Return nAP
(9) Redirect Interests toward nAP
(10) Store copies of Interests for small time ($t_s$)
(12) if $\partial$:
(13) Redirect Interests again based on nAP prefix information in $\partial$
(14) Discard the stored Interests packets
(15) else:
(16) Do nothing
(17) Discard the stored packets after time $t_s$

Figure 6: Pseudocode for new access point selection.

the new access point in our simulation. Because we have considered free-space propagation model and same transmit and receive power at all nodes, however in a realistic scenario having dynamic path losses, one can also consider other parameters to improve the accuracy. The prediction accuracy can further be improved by leveraging machine learning techniques to process the history data of mobility and predict the possible handover. The performance of prediction can also be enhanced by considering various other parameters in prediction decision such as received signal strength from each access point (AP) and the number of available channels with each AP. Considering the other parameters may improve the selection of nAP at a cost of network overhead that will be created by the routing updates (to share RSS and channels information) between the potential future new access points (e.g., nAP1 and nAP2 in Figure 7) and oAP as shown in Figure 7. Subsequently, predicting the nAP of the producer, the oAP continuously checks the reachability of the producer. The Interests are delivered to the producer as long as it is reachable, and the producer satisfies each Interest with the required content. The oAP redirects the Interest packets to nAP when the producer becomes unreachable (Step 4 in Figure 4). The oAP also stores one copy of each pending Interest for a small time $t_s$ to redirect it later, in case, the nAP prediction goes wrong. Here $t_s$ is set a little more than the RTT in our simulation to ensure the time required for nAP to notify the oAP after the handover. After the handover, if the nAP does not receive an Interest packet, it will notify its new reachable prefix to the oAP via a broadcast. If the oAP does not receive any notification from the nAP, it will assume that the nAP was predicted right and the Interest packets have successfully been redirected to it, or it will assume that the producer has completely disconnected. In both cases, the stored Interest packets will be removed after time $t_s$. The pseudocode for nAP selection and redirecting the Interests towards it is given in Figure 6. The nAP is anticipated to have producer connected to it after successful handover at layer 2. We have set the layer 2 handover delay ($L_2$) to 100 ms. Layer 2 handover delay is the time it takes by a node to disassociate itself from one AP (e.g., oAP in Figure 7) and establish a connection with another AP (e.g., nAP1 in Figure 7).

*3.2.2. Interest Redirection to nAP.* The redirected Interests are forwarded in a separate Interest packet type called INTEREST_RED. INTEREST_RED packet changes the content hierarchical name from .../oAP/ale1 to .../nAP/ale1 (since the oAP has predicted the nAP) and is forwarded based on the FIB entry corresponding to this new name. The intermediate NDN routers upon receiving the redirected Interest update the previous PIT entry for that name prefix and also notify the FIB about the changes for upcoming Interests. The intermediate routers then forward the INTEREST_RED packets according to the longest prefix matching based on FIB towards the nAP. The INTEREST_RED packets after reaching the nAP are buffered for some time and then forwarded to the producer immediately after the handover completes, and its connection establishes with the nAP. If the nAP prediction goes wrong, and the producer at the new location does not receive the Interest packets, it publishes its new name prefix by broadcasting it. The oAP after receiving this new name prefix assumes that the nAP is predicted wrong and forwards the Interests to this correct new location of the producer according to the longest prefix matching based on FIB. In this method, the pre-handover Interest forwarding to the nAP highly reduces the handover latency and overall RTT from consumer to the producer. The Interest packets path to the producer before handover is Consumer ⟶ AP ⟶ Rtr1 ⟶ Rtr2 ⟶ oAP ⟶ Producer (Figure 4). The Data packets follow the reverse path of the Interest packets. The path for the Interest packets that were in transit before updating the FIBs of the intermediate routers after handover is Consumer ⟶ AP ⟶ Rtr1 ⟶ Rtr2 ⟶ oAP ⟶ Rtr 2 ⟶ nAP ⟶ Producer. The Data packets path after handover will always be Producer ⟶ nAP ⟶ Rtr2 ⟶ Rtr1 ⟶ AP ⟶ Consumer. The scheme does not suffer from path stretch problem. Only the Interest packets that were sent before forwarding the redirected Interests suffer from path stretch.

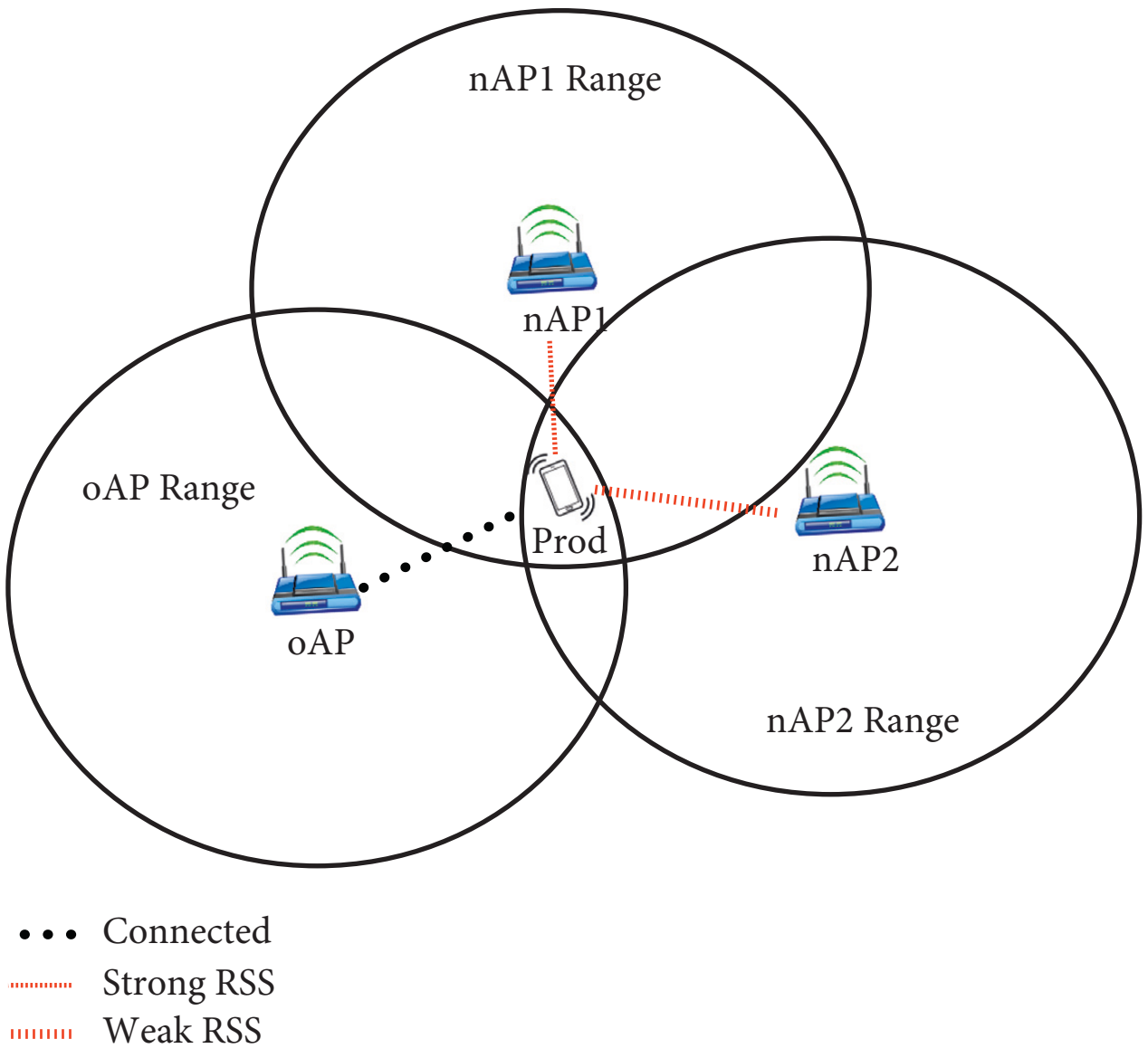

Figure 7: Access point selection.

## 4. Results and Discussion

We have simulated the following fat-tree backhaul network that resembles a real ISP topology as shown in Figure 8 using the NS-2 simulator. The scenario consists of 4 mobile producers and two stationary consumers, where each consumer is continuously sending requests for data to the two producers for the entire duration of the simulation. We have repeated the simulation 100 times, and the results are averaged over many runs. The producers randomly move, and handover can occur to any adjacent cell. We have 16 cells with full wireless coverage (WiFi 802.11n), where the producers and consumers are located in different cells initially. The 2D view of cells and the initial position of the nodes are shown in the figure. In the proposed location prediction-based mobility management methodology, the node's future expected location is predicted before finding the nAP. The oAP does not drop the incoming Interests during the handover time. Instead, oAP immediately forwards the Interests to nAP. Here RTT/2 ( ≈25 ms) (delay experienced by INTEREST_RED from oAP to nAP) is completely eliminated by overlapping it with Layer 2 handover delay ($L_2$) of 100 ms as described in Figure 9.

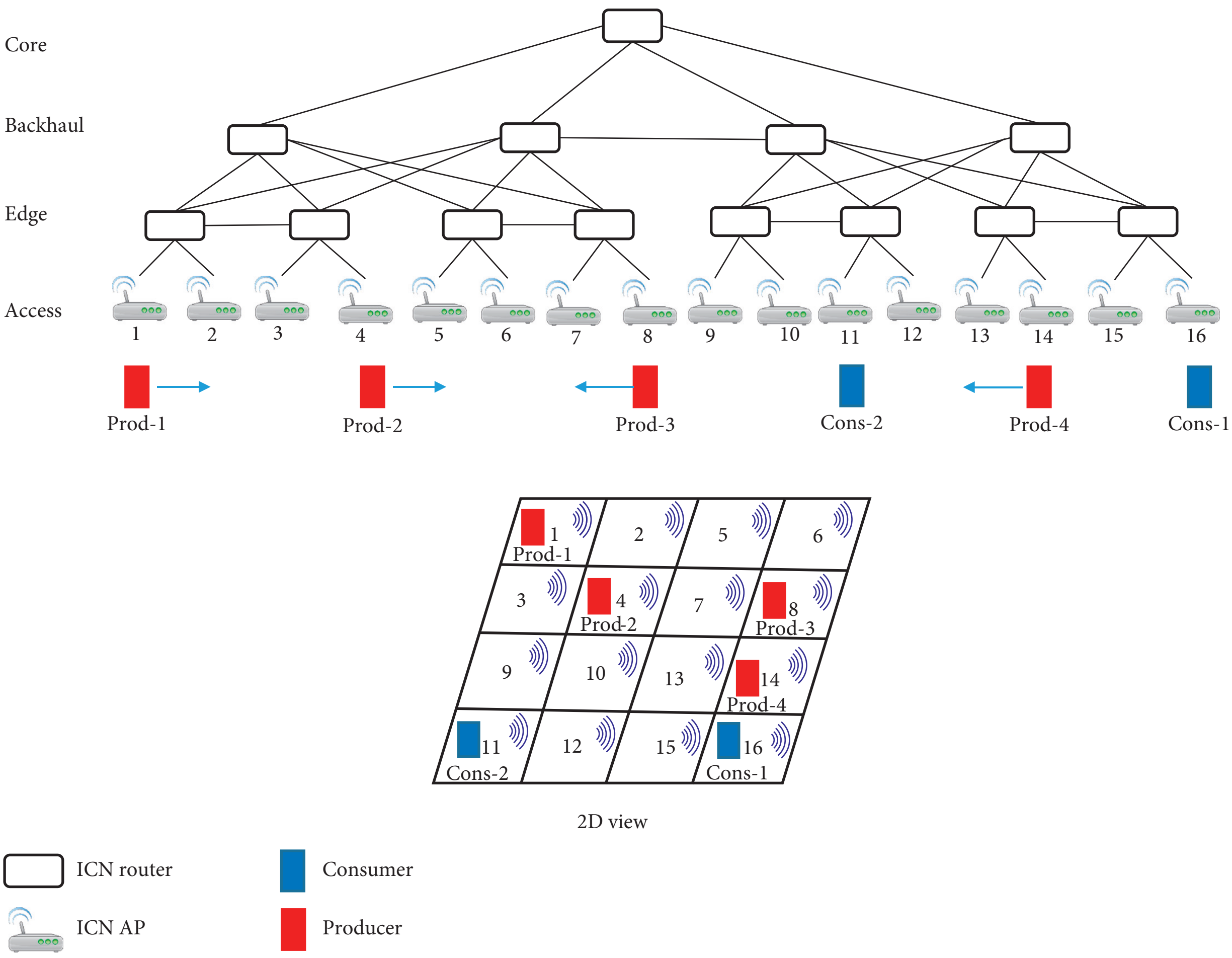

FIGURE 8: Fat-tree backhaul network topology.

It is clear that the Interest packets sent during the ($L_2$) handover delay time are delivered to the producer immediately after 100 ms. The Interest packets are not waiting for any notification from producer or network after the handover. The following equation gives the total handover latency in this scheme:

$$\begin{aligned} H_p &= L_2 + \alpha, \\ H_p &\approx L_2, \end{aligned} \tag{3}$$

where $\alpha$ is very small and covers the propagation time of Interests from nAP and producer after the handover and a little delay in predicting the actual handover incident. $H_p$ is the total handover latency, and $L_2$ is the layer 2 handover delay. The equation $H_p = L_2$ is the optimal mobility management at the network layer as it eliminates the handover delay at the network layer. The handover latency we obtained in our simulation is 109 ms for location prediction-based methodology during accurate prediction as compared to 134 ms in zone flooding and 150 ms in Interest forwarding [12] as shown in Figure 10(a) and 10(b), respectively. The better performance is because the Interest packets are forwarded to the new location (nAP) before the handover completes as the nAP is already predicted. In the Interest forwarding scheme, after the producer attached to the nAP, it sends an update message to oAP, and the oAP then redirects the following Interests towards the producer at the new location, which causes the extra latency of RTT as shown in Table 1. In zone flooding scheme, the APs are divided into zones. The consumers start multicasting the Interest to all the APs in the zone after the handover occurs. The extra latency RTT/2 is caused by sending handover notification to consumers by the producer. The AP to which the producer is connected replies with the contents, while the remaining APs ignore the received multicast.

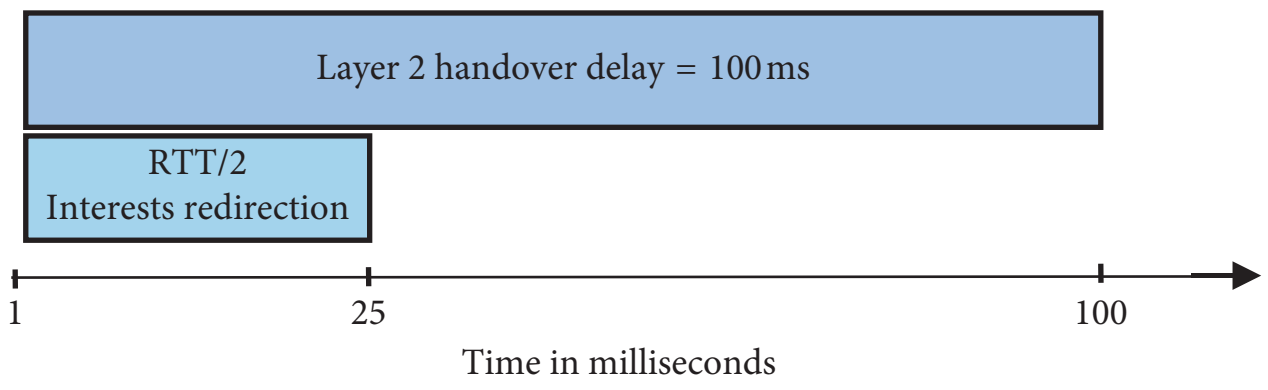

FIGURE 9: Overlapping of $L_2$ handover delay and Interest redirection delay.

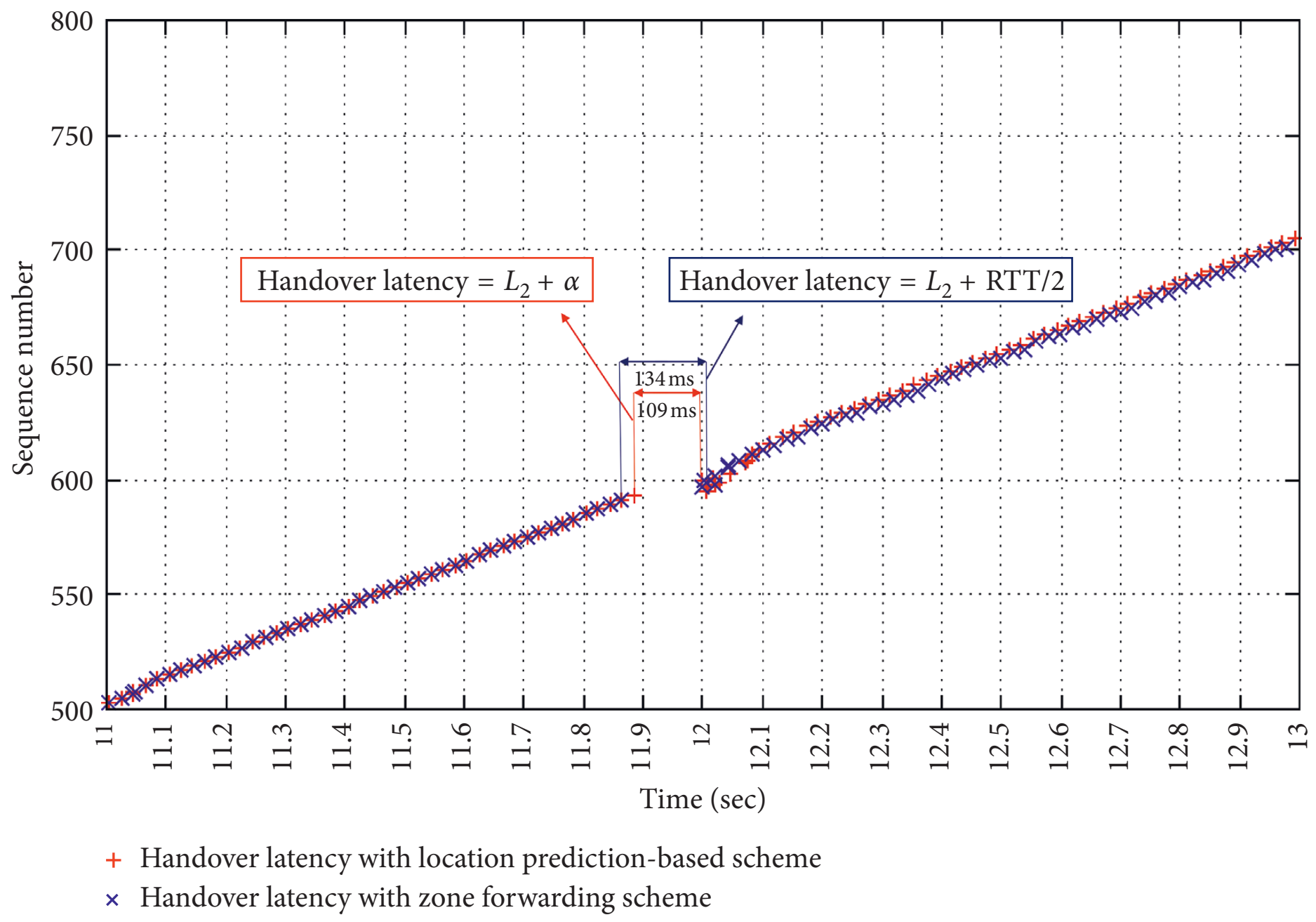

(a)

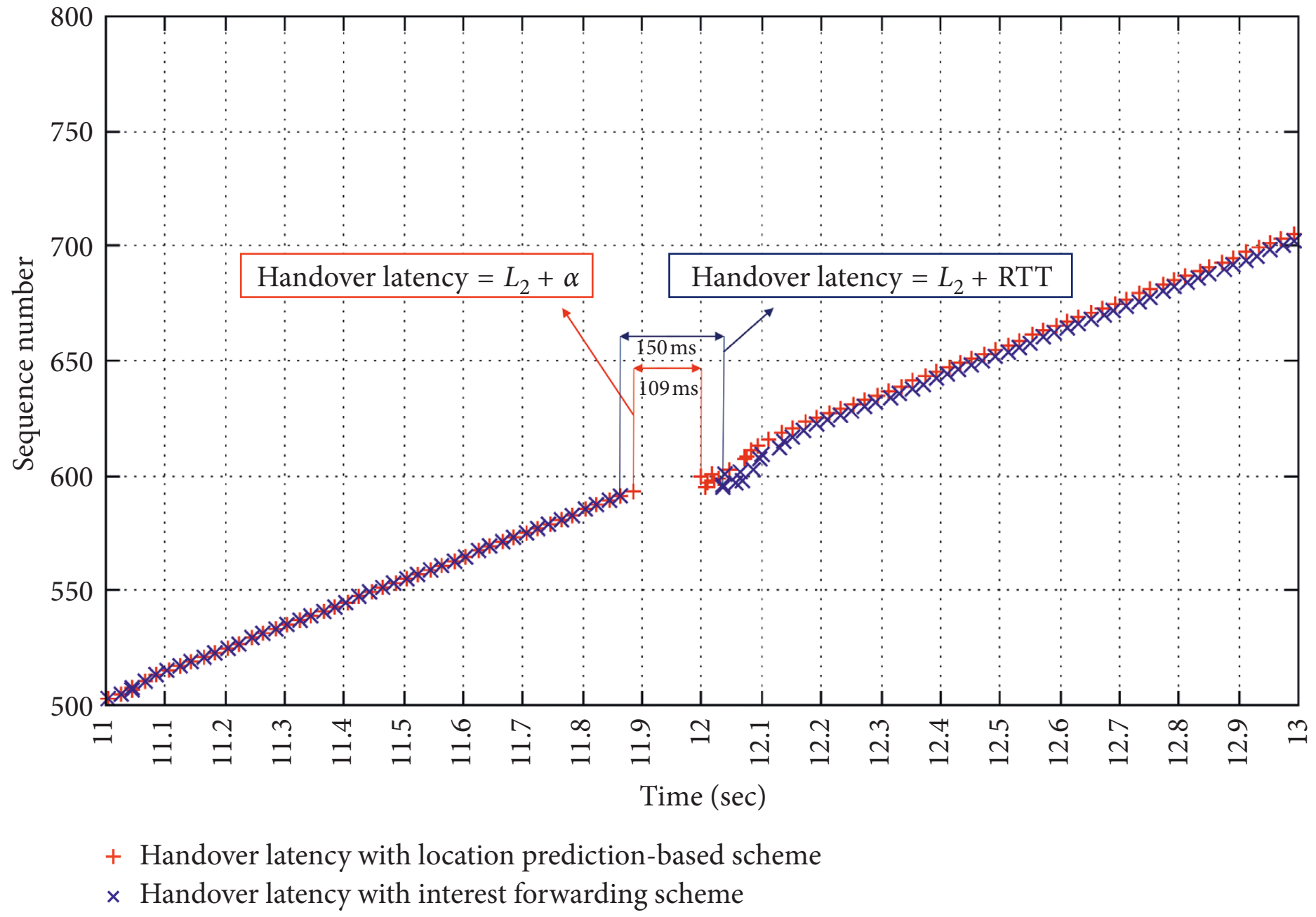

(b)

Figure 10: Handover latency in (a) location prediction and zone flooding scheme and (b) location prediction and interest forwarding scheme.

We also find the total RTT between consumer and producer. The total RTT experienced by the packets sent exactly at the time of handover ($\mathrm{RTT_h}$) in the location prediction based scheme is given by the equation below and shown by the peak value in Figure 11. RTT is defined as the total time it takes to send an Interest packet and receive the corresponding Data packet:

$$\mathrm{RTT_h} = L_2 + \mathrm{RTT_t}, \tag{4}$$

Table 1: Handover latency comparison.

| Scheme | Handover latency | Remarks |
|---|---|---|
| Interest forwarding | $H_p \approx L_2 + \text{RTT}$ | RTT between producer old location and new location is added which varies with topology and other network parameters |
| Zone flooding | $H_p \approx L_2 + \text{RTT}/2$ | RTT/2 is added here which also varies as RTT varies |
| Location prediction based (proposed) | $H_p \approx L_2$ | RTT is completely eliminated with our proactive approach (Figure 9) |

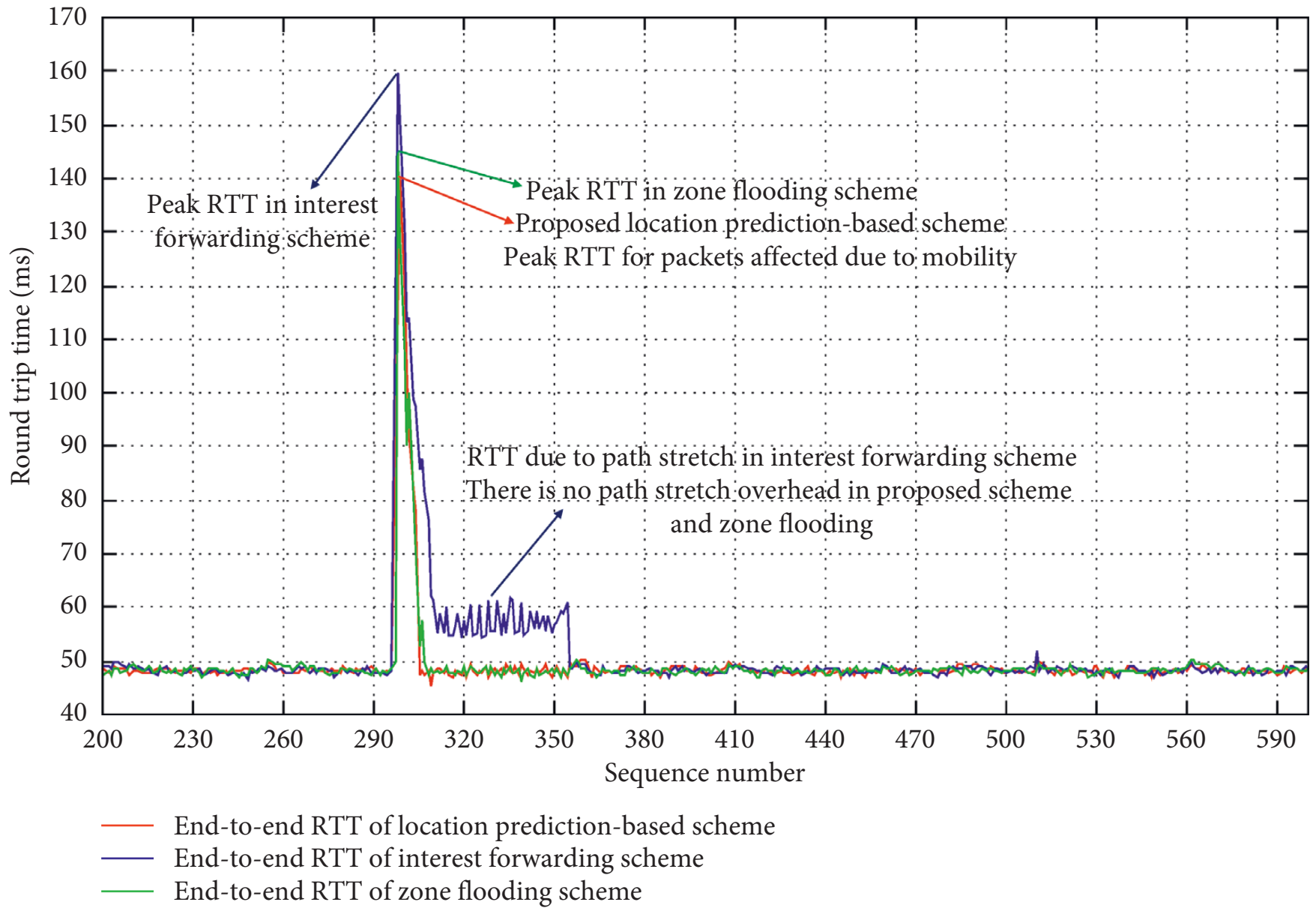

Figure 11: Round trip time comparison of the three mobility management schemes.

where $\text{RTT}_t$ is the average round trip time between consumer and producer before handover. The RTT during handover is highly reduced to 139 ms, and RTT in Interest forwarding scheme is 160 ms, while in zone flooding, RTT is 144 ms, almost similar to our scheme. However, the zone flooding scheme creates a very high network overhead as it is based on sending multiple copies of Interests to all APs in a zone. Therefore the zone flooding scheme is not scalable. On the contrary, the location prediction-based scheme creates no network overhead. RTT due to path stretch in the Interest forwarding scheme is shown in Figure 11, while the location prediction-based scheme does not suffer from the path stretch problem.

Figures 12(a) and 12(b) show average handover latency in different schemes with varying percentage of accuracy in prediction. Since the prediction cannot be 100% right, Figure 12(a) shows that with a very low prediction accuracy of 50%, location prediction based method substantially reduces the average handover latency, yet keeping excellent content to the Interest ratio as shown in Figure 12(b). RTT and handover latency for zone flooding are quite small, but its content to the Interest ratio is also very small because of its multicasting nature, and it cannot be regarded as the best solution. Moreover, our proposed approach significantly reduces average handover latency, RTT, and maintains an outstanding content to Interest ratio as evident from the simulation results. On the contrary, Interest forwarding scheme has a good Content to Interest ratio, but it has big RTT, handover latency, and suffers from path stretch.

## 5. Conclusion and Future Work

In NDN, the producer mobility is one of the strenuous and challenging tasks during real-time streaming. The handover latency constitutes the delay caused by dissociation and reassociation of a mobile node to oAP and nAP, respectively ($L_2$ delay), and the delay caused by network convergence time. In this work, we have exploited location prediction for producer mobility management to show its impact on the design of such mobility management techniques. The simulation results have shown that using location prediction techniques for mobility management can significantly reduce the total handover latency to approximately equal to layer 2 handover delay by eliminating the delay caused by network convergence. With a minimal prediction accuracy of 50%, the methodology tends to perform better, which shows the benefits of using location prediction for producer mobility management. It also reduced end-to-end RTT without creating any network overhead. The results of this mobility management approach may turn out as the base knowledge for the design of a sophisticated location prediction technique for mobility management. As of future work, we plan to design an efficient location prediction

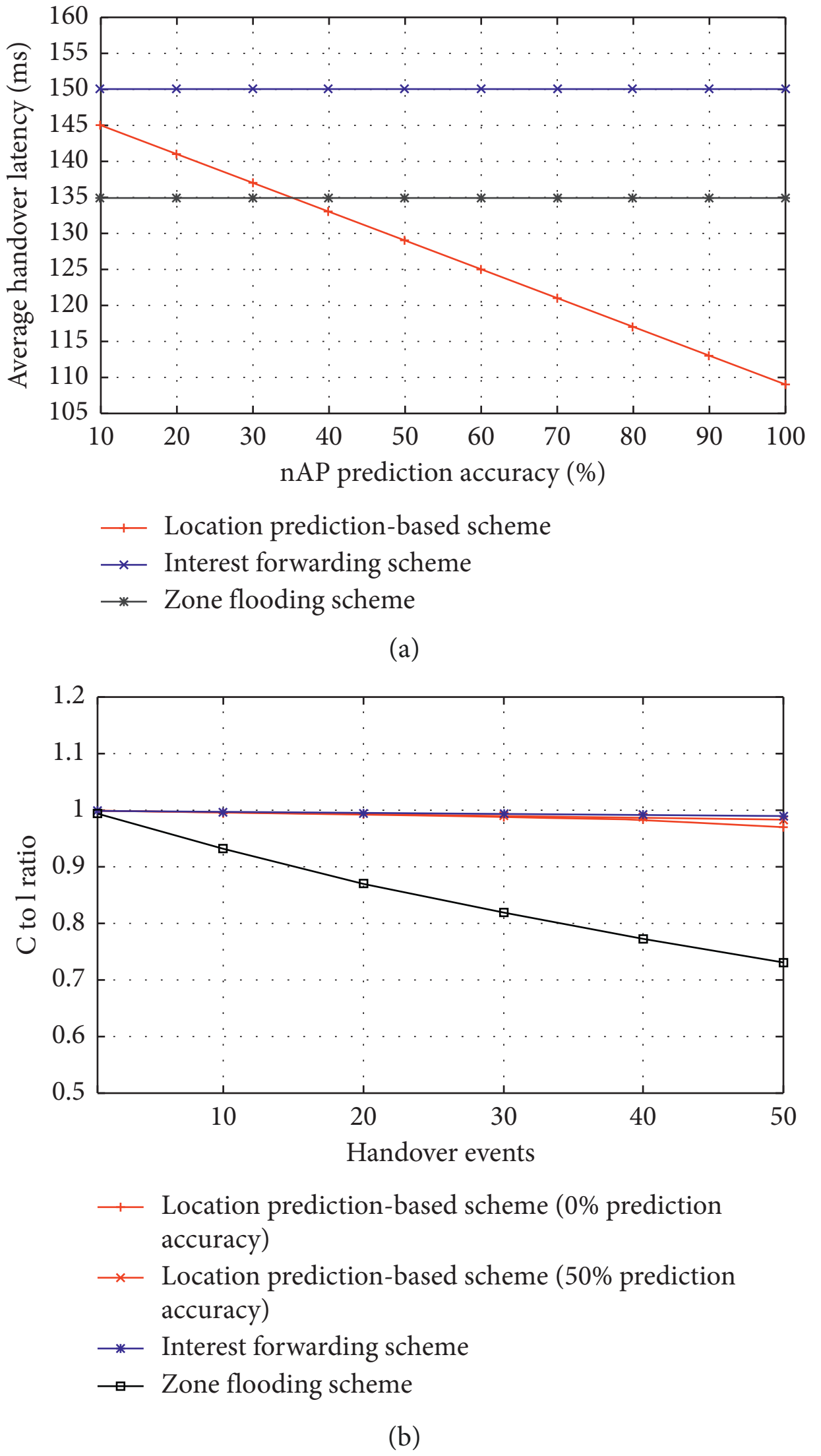

Figure 12: (a) Average handover latency. (b) Content to interest ratio.

technique and apply it in our proposed methodology for mobility management in NDN.

## Data Availability

No data were used to support this work. Moreover, the simulation code is available from the authors upon request.

## Conflicts of Interest

The authors declare that they have no conflicts of interest.

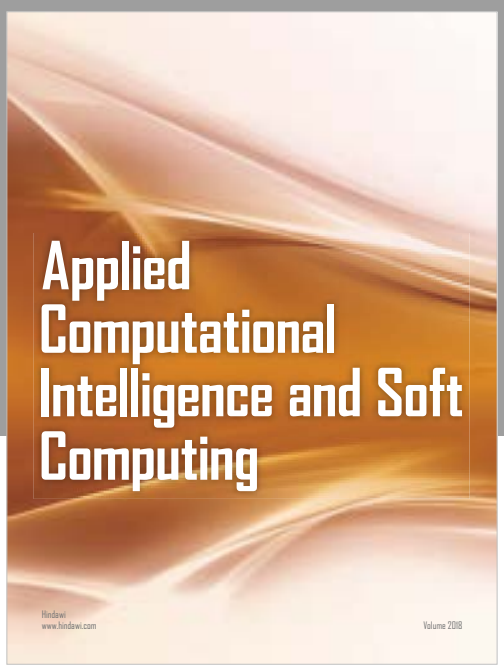
Applied Computational Intelligence and Soft Computing
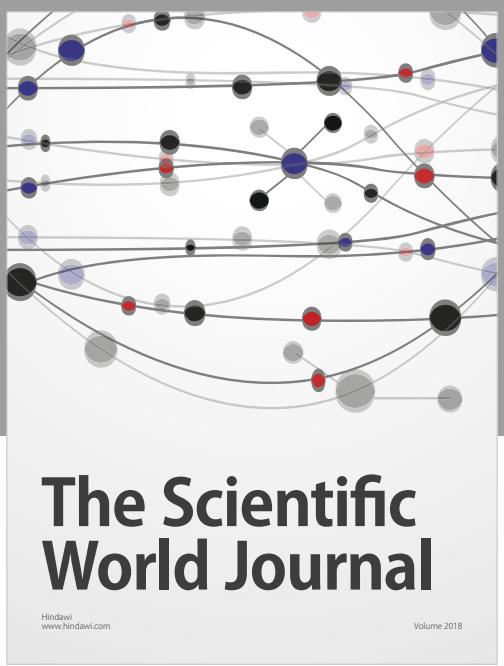
The Scientific World Journal
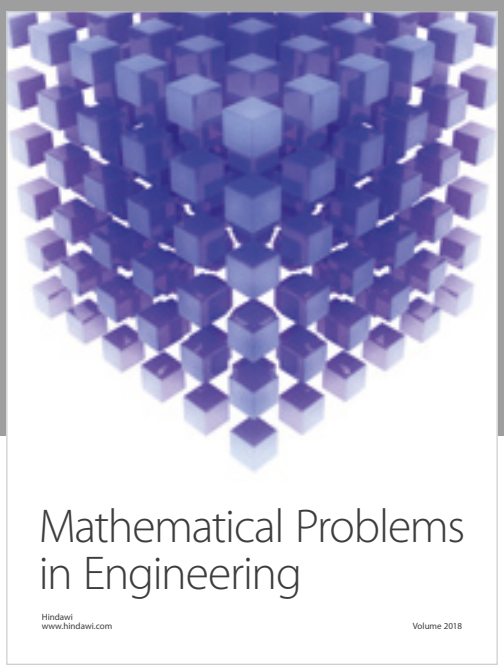
Mathematical Problems in Engineering
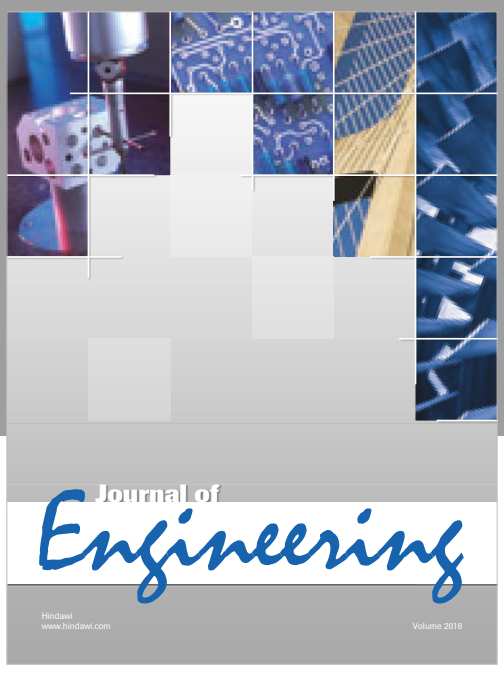
Journal of Engineering
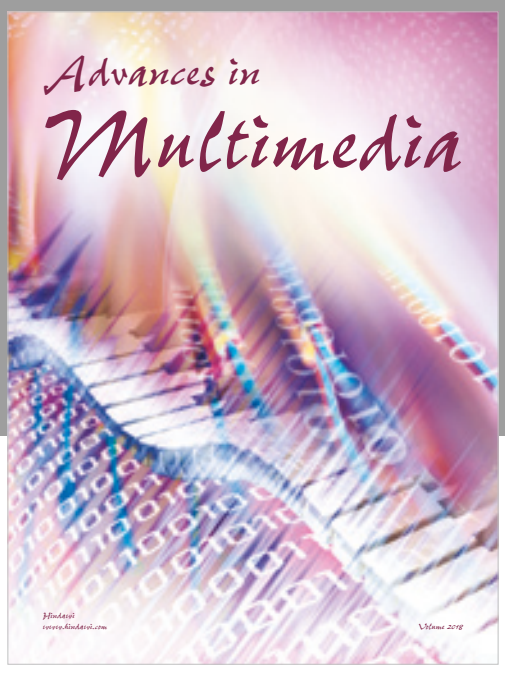
Advances in Multimedia
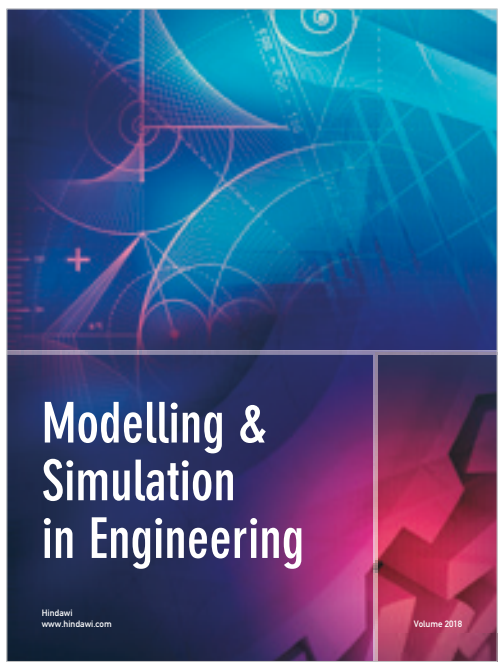
Modelling & Simulation in Engineering

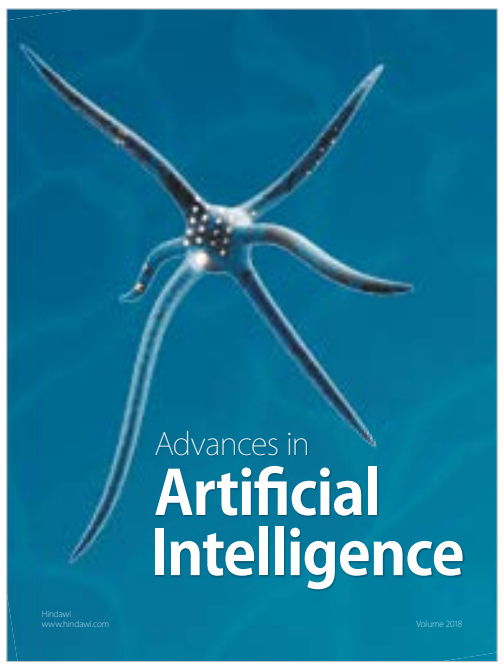
Advances in
Artificial Intelligence
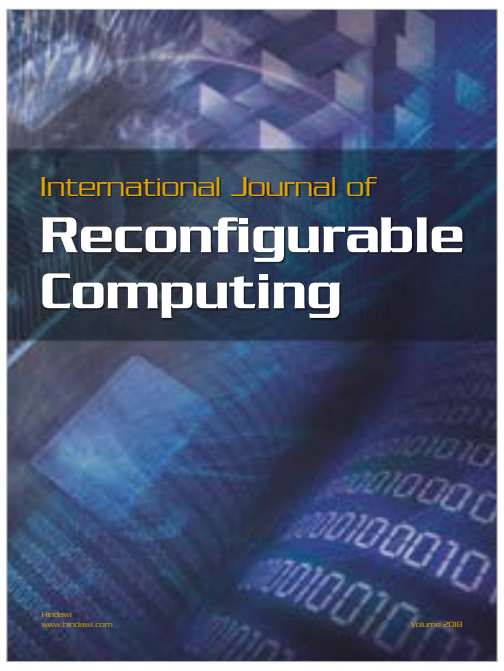
International Journal of
Reconfigurable Computing
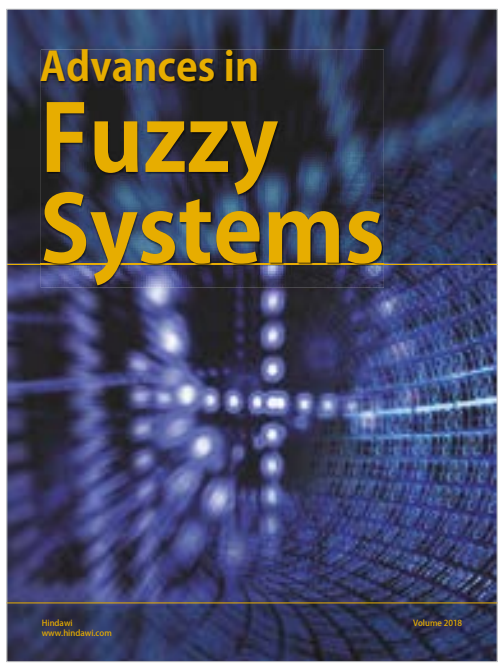
Advances in
Fuzzy Systems
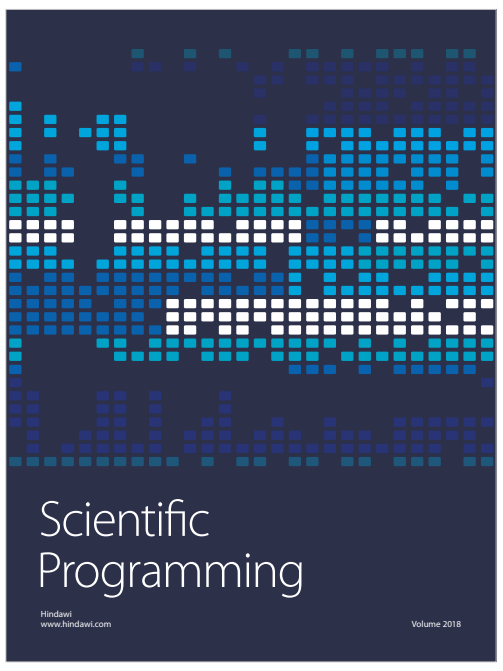
Scientific Programming
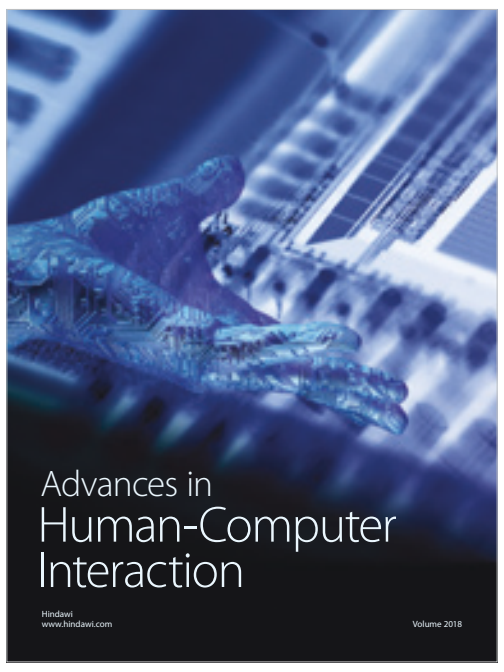
Advances in
Human-Computer Interaction
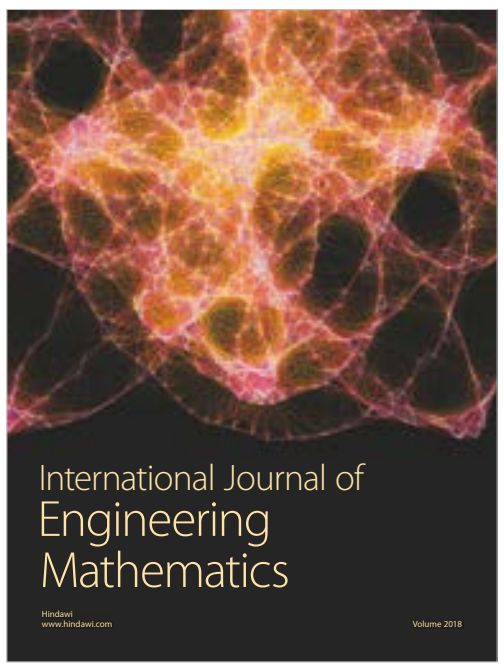
International Journal of
Engineering Mathematics
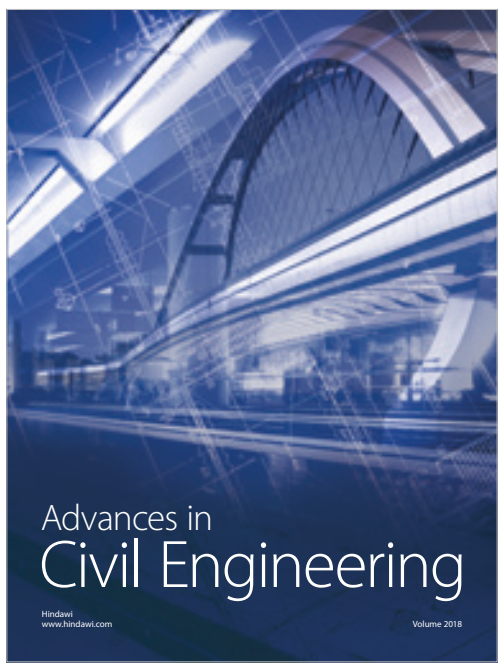
Advances in
Civil Engineering
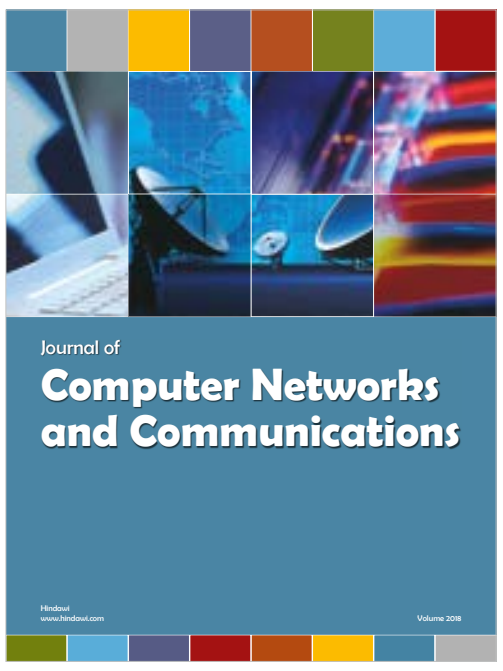
Journal of
Computer Networks and Communications
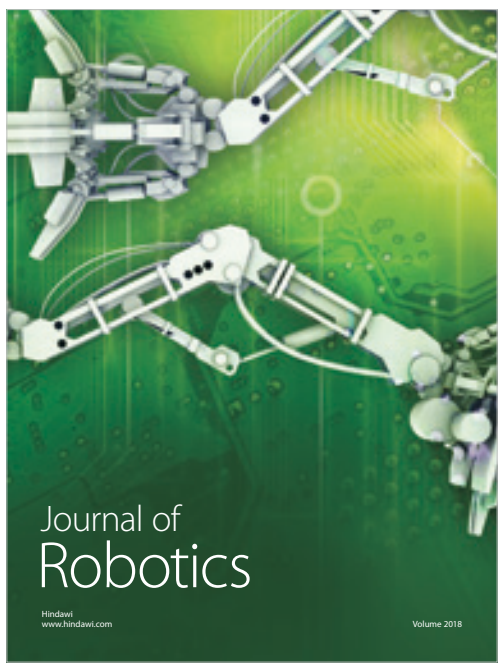
Journal of
Robotics
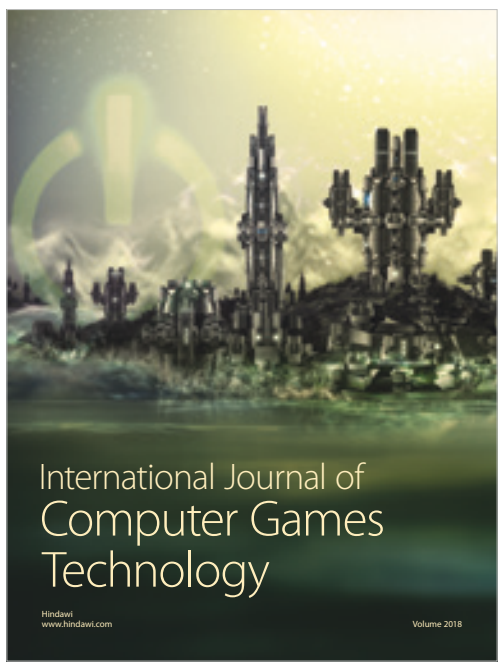
International Journal of
Computer Games Technology
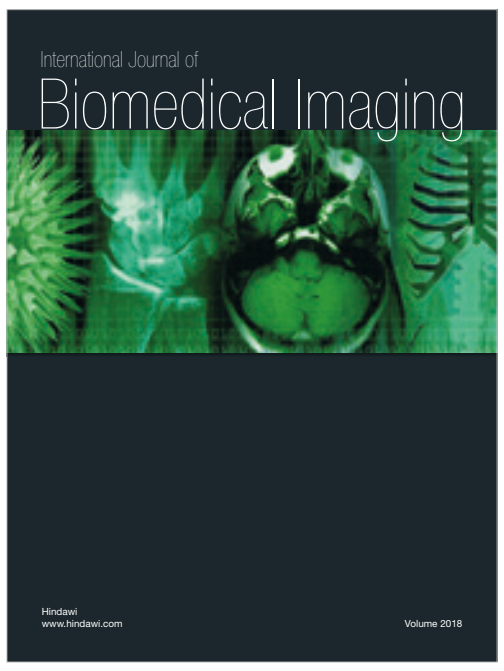
International Journal of
Biomedical Imaging
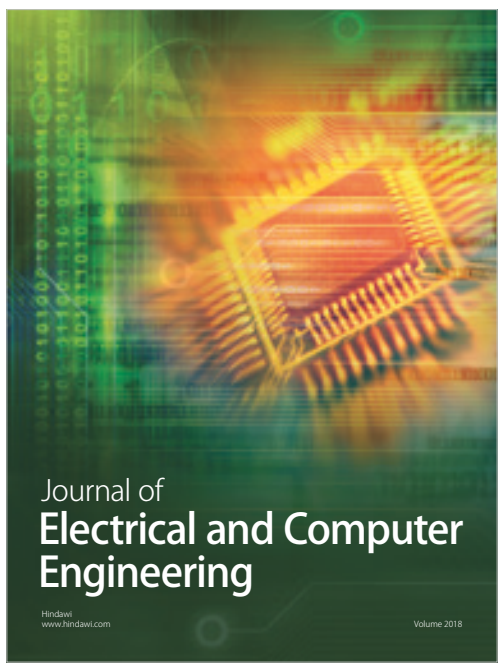
Journal of
Electrical and Computer Engineering
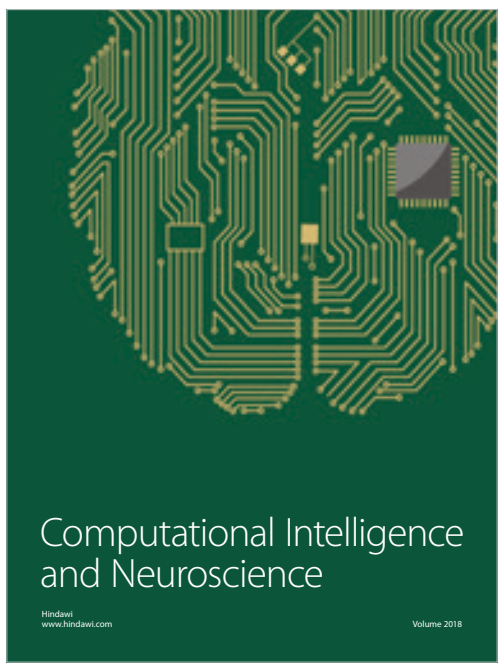
Computational Intelligence and Neuroscience